\documentclass[12pt]{article}

\author{Vladimir A. Petrov \footnote{e-mail:Vladimir.Petrov@ihep.ru} }

\title{On the `` Froissaron-Maximal Odderon" Model.}
\date{}

\begin{document}

\maketitle
A. A. Logunov Institute for High Energy Physics, NRC KI,
Protvino, RF

\begin{abstract}
We analyse the basic premises of the `` Froissaron-Maximal Odderon" (FMO) model which was claimed to be ``the only existing model which describes the totality of experimental data ". It is shown that the FMO model suffers from serious theoretical flaws while its quality of the data description is such that the probability that it describes the selected set of data is not satisfactory enough.
\end{abstract}

\section{Introduction}

In 1973 the Odderon concept was pushed forward \cite{Nic} which embodies a hypothesis that the leading C-even agent dominating high energy behaviour of the strong interaction cross-sections, the famous Pomeron, may have a C-odd counterpart which can give non-negligible contributions at high energies in contrast  to secondary Reggeons $ \rho,\omega $ etc rapidly  dying off with the energy growth. Generally, it is but natural that such a counterpart to the Pomeron should exist if the very Pomeron exists as a predominantly gluonic exchange.

Another issue is which are the nature and characteristic properties of this entity.
The simplest option is a simple pole in the j-plane, then nobody can forbid several or even infinite number of simple poles, some author admit double and even triple poles,
the use of the Regge-eikonal framework allow to deal with high intercepts of these Regge poles without violation of the sacred upper bounds, finally you can just invent a C-odd contribution to the scattering amplitude at your will and for your purposes. The mentioned upper bounds restrict the properties of one or another Odderon choice from almost imperceptible at high energies to the maximally allowed.

One of the motivations of ``maximality" of the Odderon was an extension of the old Chew-Frautschi imperative  of "maximum strength" for strong interactions at high energies \cite{Che}  from the asymptotic constancy of the total cross-sections to the functional saturation of the Froissart bound implying their $ \sim \ln ^{2} s $ behaviour but, more than that, the similar saturation of the upper bound for the differences of "C-conjugated ``processes (e.g. for $ \Delta\sigma = \sigma_{tot}^{\bar{p}p} - \sigma_{tot}^{pp}$) which is bounded above ( \textit{"modulo modulus"}) by $ \sim ln s $.
Thus, the ``maximal Odderon" hypothesis implies the violation of the Pomeranchuk theorem (in the sense of differences)which asks for  $ \Delta\sigma\rightarrow 0 $ and, in a more general context, the violation of the Gribov "principle of asymptotic universality" \cite{Gri} extended to the case of rising cross-sections. 

From a (maybe a bit simplistic) physical viewpoint the idea of the maximal Odderon implying, in particular, that the difference between the proton-proton and anti proton-proton interactions becomes more and more noticeable ($ \Delta\sigma = \sigma_{tot}^{\bar{p}p} - \sigma_{tot}^{pp} \sim \ln s $ ),  predicts at the same time that the average distance (impact parameter) between the colliding particles (``interaction range") increases\footnote{We mean the well-known increase of the forward slope $ B(s)\approx \langle b^{2} \rangle/2 $ with the energy growth which is also featured in ``maximal Odderon".}.
In very simple words it means that colliding particles - according to the ``maximal Odderon" model-  appear to see the difference in their inner structure (which is defined by the energy independent "valence core" \cite{Okor} ) the better, the farther they are from each other.

But no matter how strange such behaviour would seem to us, such an unorthodox option  realized in the long-term efforts by B. Nicolescu \cite{Nic}, \cite{MaNi} , seems to be fairly conceivable in a formal sense and even shows up a certain elegant symmetry for forward amplitudes $ T(s,t=0)= T(s,0) $ because it implies that ``at high energies"
\begin{equation}
T^{+}(s,0) = iA\cdot ln^{2}(s\cdot exp(-i\pi/2)),
\end{equation}
\begin{equation}
T^{-}(s,0) = B\cdot ln^{2}(s\cdot exp(-i\pi/2))
\end{equation}
where 
\[T^{\pm }(s,0) = \frac{1}{2}[T^{\bar{p}p}(s,0) \pm T^{pp}(s,0)] .\]

Let us note that not every pair of ``C-conjugated" processes  can be associated with the Odderon exchange\footnote{The term ``exchange" is used in a wide sense meaning only the general quantum number and energy-momentum exchange irrelevant to a concrete mechanism.}. 
E.g., the pair of processes
$ \pi^{\pm} p \rightarrow \pi^{\pm} p $ is a counter example. We, however, will deal with 
(anti)proton-proton scattering which is under active discussion in relation with a rich experimental material accumulated by now by the collaborations TOTEM and, partly, by ALFA(ATLAS).
Beyond the elastic scattering processes other options for searches for the Odderon manifestations are actively discussed in the literature\footnote{Interesting proposals to observe possible Odderon signals in the central meson production processes were suggested, for instance, in Refs. \cite{Kho} . }.

As to the model realization of the very Odderon contribution they are too abundant - both in pure theory and phenomenology - to make a review of them in the present article narrowly devoted to a thorough analysis of a chosen approach.

From the very beginning of the maximal Odderon idea it was related with expectations of a cardinal, both qualitative and quantitative, change of mutual relation between the total $\bar{p}p  $ and $ pp $ cross-sections, viz., the cross-over 
changing $\bar{p}p $ dominance to the $ pp $ one. With time the predicted energy where such an event should occur moved from $ \mathcal{O}(20) GeV $ to the recent estimate near $ 300  $ GeV.
In the absence of simultaneous measurements of  $\bar{p}p $ and  $ pp $ it is quite difficult to observe such a cross-over: the difference of order $ \mathcal{O}(1 mb) $ is expected at c.m.s. energies no less than $ 100 $ TeV, far beyond any realistic plans.
Nonetheless, there are other features of the maximal Odderon which could be tentatively caught with now existing means. For instance, a latest embodiment of the maximal Odderon doctrine, the `` Froissaron- Maximal Odderon" model  (FMO), yielded the value of the parameter $ \rho \doteq ReT^{pp}(s,0)/Im T^{pp}(s,0) = \mathcal{O} (0.1)$ almost exactly coinciding with its value published by the TOTEM Collaboration \cite{Ant}\footnote{Note in passing that the very method of deriving the value of $ \rho $ from the data in the region of Coulomb-nuclear interference, used in \cite{Ant} , was disputed in the works \cite{Ppe} .} . 

This coincidence gave an impetus for the latest modification of the $ t\neq 0 $  extention of the maximal Odderon \cite{NicMa}  with a consequent conclusion that
`` the Froissaron-Maximal Odderon (FMO) approach
is the only existing model which describes the totality of experimental data (including the TOTEM results) in a wide range of energies and momentum transfers" \cite{Ni}.

Thus, in view of such a high-flown statement, this is but natural to try to understand in more detail the basic features and premises of the FMO. 

To the conceptual issues we also add 
a statistical estimate related to Tables 1, 2 and 3 from \cite{NicMa} in terms of p-values. 

 Results of such an inspection - to the best of the author imperfect understanding - are exhibited in Sections 2 and 3.

\section{Analysis of the FMO model}
 The model in question consists as described in  \cite{NicMa} of the following  components:
 
 1. Standard secondary C-even ($ R_{+} $) and C-odd ($ R_{-} $)Reggeons with low intercepts $ < 1 $;
 
 2. C-even (`` simple Pomeron") and C-odd Regge poles (``simple Odderon") with intercepts $ \alpha_{P, \:O}(0)=1 $ .
 
 3.  C-even and C-odd  ``maximal elements" embodied in specially designed functions
 $ \varphi ^{\pm} (\omega=j-1,t) $  which are to provide the mentioned above ``maximising" properties of $ \Delta\sigma $
 and $ \sigma_{tot} $.
 
In addition ``interference" terms are assumed which are composed pairwise of the above mentioned elements 1-3.
 
 Below we concentrate on functions from point 3 above as they  constitute the backbone of the approach. 
 The amplitude $ T(s,t)$ is related to signatured functions $ \varphi ^{\pm} (\omega,t) $ as follows ( as implies Eq.(18) from Ref.\cite{NicMa} )
 \begin{equation}
 T(s,t)= z_{t}\int_{C} \frac{d\omega}{2i\pi}e^{\omega \varsigma} \varphi ^{+} (\omega,t)+ z_{t}\int_{C} \frac{d\omega}{2i\pi}e^{\omega \varsigma} \varphi ^{-} (\omega,t)
 \end{equation}
 where 
 \[z_{t}= \frac{s-u}{4m^{2}-t} = \frac{2s}{4m^{2}-t} - 1;\; \zeta = \ln z_{t} - i\pi/2 \]
 and the integration contour $ C $ is $ \omega = \omega^{'} +i \omega^{''} $ with $ \omega^{'} $ lying to the right of the rightmost singularity of $ \varphi ^{\pm} (\omega,t) $ in the complex $ \omega $- plane and $ - \infty <  \omega^{''} < + \infty $.
 The functions $ \varphi ^{\pm} (\omega,t) $  are assumed to be (Eq.(23) in  \cite{NicMa} )

 \begin{equation}
 \varphi ^{\pm} (\omega,t) =  {i\choose -1} \frac{\beta^{\pm}(\omega, t) }{(\omega^{2}- r_{\pm}^{2}t)^{3/2}} .
 \end{equation}
 
 The reason for such a choice: it should provide the asymptotic behaviour as in Eqs.(1) and (2). 
 One  may notice that the option for $ \varphi^{+} $ formally coincides with the well known partial amplitude following from  Regge-eikonal representation with the `` supercritical" Pomeron with
 \[r_{\pm}^{2}= 4(\alpha_{\mathcal{P}}(0)-1)\alpha^{'}_{\mathcal{P}}(0)\]
 where $ \alpha_{\mathcal{P}}(t) $ is the Pomeron pole trajectory.
 We have, however, to keep in mind that in normal Regge-eikonal approach the counterpart of $ \varphi^{+} $ does not `` swallow" the factor $ 1/\cos (\pi\omega/2)  $.
 
 If we look at Eq.(3) we notice that the integral along the line $ Re\omega = \omega^{'} $
 diverges at $ \omega^{''}\rightarrow +\infty $ as $ \exp (\pi\omega^{''}/2) $.
 The fate of this integral depends critically on the properties of the function $ \beta^{\pm}(\omega, t) $ at $ \omega^{''}\rightarrow +\infty  $.
 Let us take for definiteness the amplitude $ \varphi ^{-} (\omega,t) $.
 According to Eq.(4) 
 the only source of damping the exponential growth would be the opposite trend of $ \beta^{\pm}(\omega, t)$ which could provide the Sommerfeld-Watson factor $ 1/sin (\pi\omega/2) $
normally doing the job but Eq.(24) in Ref.\cite{NicMa}, according to which 
 \begin{equation}
 \phi^{-}(\omega, t) \sim \sin (\pi\omega/2)\varphi ^{-}(\omega, t)
 \end{equation}
and therefore this factor is cancelled out, deprives us of this hope. 

It remains to assume that some other source could probably resolve the problem and lead authors to the \textit{bona fide} amplitudes, e.g. the (maximal) Odderon contribution like

\[\frac{1}{z}F_{MO}(z_{t},t) = \textit{O}_{1}\zeta^{2}\frac{2J_{1}(r_{-}\tau \zeta)}{r_{-}\tau \zeta} \Phi_{\textit{O},1}^{2}(t)+...\]
(their Eq.(29)) which they successfully use for the TOTEM data fitting. Unfortunately, we could not find such instructions in the main text\footnote{Except the Appendix A which is commented a bit below}  of \cite{NicMa}.

One of the general properties the observance  of which is strongly required is the observance of the unitarity condition. Taking ,e.g., s-channel unitarity it means, in particular, that 
\[\mid \tilde{T}(s,b) \mid < 1\]
where
\[\tilde{T}(s,b) = \frac{1}{8\pi s} \int dt J_{0} (b\sqrt{-t})T(s,t)\]
is the scattering amplitude as function of the impact parameter $ b $.

In Ref.\cite{Kh}  the paper \cite{NicMa} (a) was criticized for a supposed violation of the unitarity condition. We have to notice that this criticism is hardly tenable because it refers to the Regge-eikonal formalism which is not in use in \cite{NicMa}.

Eqs.(28)-(29) from Ref.\cite{NicMa}(b) which are generalization of \cite{NicMa} (a) to non zero $ t $ allow to obtain the following asymptotic behaviour of the partial amplitude (it is enough to take the worst case $ b=0 $) :

\[\mid \tilde{T}(s, b) \mid _{s\rightarrow \infty, b =0}  \:\ \rightarrow \frac{1}{4\pi}\mid iH_{1}/r_{+}^{2} + O_{1}/r_{-}^{2} \mid  \approx  0.88\]
with parameters from Table 3 from \cite{NicMa} (b).

Even a more restrictive inequality
\[Im\tilde{T}(s, b)> \mid \tilde{T}(s, b) \mid^{2} \]
is respected.
So the FMO model does not violate the unitarity condition.

Unfortunately, this fact is not enough to recognize the FMO model as at least a passable one.

The matter is that the above mentioned  effective absence of the factor $ 1/sin (\pi\omega/2) $ leads to a striking property: at all even $ \omega $ ( all odd $ j $) the physical partial amplitudes $ \phi^{-}(j=2N+1, t)= \phi(j=2N+1, t) $ which enter the Legendre expansion of the scattering amplitude (cf Eq.(16) in Ref.\cite{NicMa}):

\begin{equation}
T(z_{t},t)= 16\pi \sum_{j=0} ^{\infty} (2j+1)P_{j}(-z_{t}) \phi(j, t)
\end{equation}
turn out to be zero as follows directly from Eqs.(4) and (5).

In order not to be unfounded, let us quote the authors of \cite{NicMa} who wrote, in particular, the following:  `` the real\footnote{The authors evidently use the word `` real" in the sense of `` true".}  physical partial amplitude...$ \phi^{-} (\omega, t) $... equals to zero at $ \omega=0 $ " (see the text around Eq.(24) of their paper).
Actually, as we have seen above ( Eq.(5)), the genuine physical partial amplitude $ \phi^{-} (\omega, t) $ vanishes not only at $ \omega=0 $ but at all even $ \omega $ (odd $ j $).
In a similar way we find that $\phi(j, t)$ vanishes at even $ j = 2N$ as well.

 At last, there is one more confusing circumstance. Appendix A of the same paper introduces new amplitudes $ \phi_{\pm} (\omega, t) $ and $ \varphi_{\pm} (\omega, t) $ instead of above mentioned $ \phi^{\pm} (\omega, t) $ and $ \varphi^{\pm}(\omega, t) $. They are happily free from the defects above. Misprint? 
 
 Alas, even if so, we find that the physical $ p $-wave amplitude in the $ t $-channel ($ \bar{p}p $ annihilation) has a singularity at $ t=0 $:
 \[\phi(j=1, t)= \phi_{\pm} (\omega=0, t) \sim (-t)^{-3/2}.\]
 So, now there is evidently `` something" C-odd which is really odd in the $ t $-channel. Not a pole but rather a branch point, kind a `` singular threshold" related to massless states. The only source of such a threshold could be unconfined gluons but the authors do not go that far. 
 A discovery of the deconfinement would be certainly a significant finding. But in this case we would have a lot of other processes where such events would reveal themselves.
However, nothing of this kind was ever reported.

\section{ Discussion and conclusions}

 \begin{flushright}
 ``\textit{Criticism, like lightning, strikes the highest peaks}."
 
 Baltasar Gracián, The Art of Worldly Wisdom. 
 \end{flushright}

\subsection{Physical content}
As we have seen, the FMO model in its main part (see p.3 in the previous Section) seems to exhibit quite unusual features. In particular, the very scattering amplitude does not have physical states in the t-channel.Probably this bobble is inherited from previous versions of the FMO model(see, for example, \cite{MaNi}). 

The fact that in spite of zero amplitudes for physical (integer ) $ j $, a nonzero amplitude can be obtained is explained by the violation of Carlson's theorem as can be seen from equation (5). 

Another option which can be found in the Appendix A to the paper seems to be free of all this mess but instead it demonstrates a  massless singularity in the $ p $-wave of the $ \bar{p}p $ scattering challenging confinement of QCD.

\subsection{Descriptive qualities}
If to ignore for a moment the theoretical flaws discussed above we could think that at least the description of the data exhibited in \cite{NicMa} and \cite{Ni}  is as outstanding as quoted in the end of Introduction which would make a due reconsideration of the theoretical premises worth the new efforts.

Unfortunately,  this is not the case. If to apply, taking use of Tables 1, 2 and 3 from \cite{NicMa}, the p-value (probability that the model describes the selected set of data) criterion ( see, e.g. \cite{PDG})   then we see that Table 1 (FMO without Coulomb)
corresponds to a p-value  of $ p = 4.7\cdot 10^{-70} $. This could mean that omission of the Coulomb contribution leads to practically zero probability with a greater than 8 sigma standard deviation between the FMO(without Coulomb) and the data.

However, Table 2 shows that inclusion of the Coulomb interactions provides a p-value of 
$ p = 5.4\cdot 10^{-84} $ while the use of the "two-sided" p-value gives a bit better estimate $ p = 1.09\cdot 10^{-83} $ which, certainly, is also far from satisfactory\footnote{We are sorry to add that the values of some parameters in
Ref.\cite{NicMa} are simply missing or evidently misprinted.}.

Thus, the FMO model as presented in \cite{NicMa} and \cite{Ni} does provide a description of a large set of data with a very large number of parameters (36 and 24, respectively, counting the number entries in Table 3) but it does not provide a statistically acceptable representation of the data, hence it cannot be used to claim evidence for the Odderon at least based on good quality data description.

In technical terms, we are, of course, aware of the enormous difficulties in describing the huge array of data (3226 points on the whole!)  that were used in Ref.\cite{NicMa}   and if we were only worried about the chi squared then $ \kappa^{2}/ndf =1.6 $ would look quite good ( leaving aside p-value criteria) and in this respect the work should be recognized unique in the present literature.

\subsection{Summa summarum}

\textit{Therefore the statement, made in \cite{Ant} on the basis of comparison with the FMO model and the `` extended Durham model"\cite{Dur}  that the TOTEM measurements at 13 TeV supposedly discovered the evidence of a `` 3-gluon compound state" in the t-channel  does not seem well justified in the part concerning the FMO model.}\footnote{As to the `` extended Durham model" \cite{Dur} , we could not find in corresponding publications the information on the $ \chi^{2} $ etc which would allow us to estimate its statistical reliability.} 

 Of course, this does not  devalue at all
the very experimental data obtained by the TOTEM Collaboration, but only indicates problems with some of their possible theoretical interpretations.

The Odderon was the subject of other theoretical papers as well ( see for instance \cite{Sel}) but this paper was not designed as a review, so we limited ourselves by papers preferred by the TOTEM collaboration.

Evidently, in general, it is quite problematic to extract an unambiguous information on the Odderon properties relying only on $ pp $ scattering data, without an `` accompanying" data from $ \bar{p}p $ interactions at the same energy.

In this respect, we have to mention the recent attempt to `` renormalize" (with the help of an a novel `` transfer" technique ) the $ pp $ data from TOTEM at 2.76 TeV to 1.96 TeV which allows to make a comparison with the $ \bar{p}p $ data from D0 \cite{Ro}.  Viz., with such new $ pp $ `` data" we can compare the ratio  $ [d\sigma^{\bar{p}p}/dt|_{t=t_{break}^{\bar{p}p}}]/[d\sigma^{pp}/dt|_{t=t_{dip}^{pp}}] $ at $ \sqrt{s}$ = 53 Gev and 1.96 TeV. The ratio drops from $ \approx 3.6 $ at 53 GeV as can be seen in Ref.\cite{Bre} (though with an overall uncertainty $ \pm 30 \% $ ) to $ \approx 1.5 $ at 1.96 TeV. If to assume that secondary C-odd trajectories are already negligibly small at the ISR then one could argue that the Odderon effect
($ d\sigma^{\bar{p}p}/d\sigma^{pp} > 1 $ ) in the vicinity of $ pp $ dip seemingly dies off (though not so fast) with the energy growth.

\textit{This shows again that the right place to search for the Odderon effects is not so much the `` forward" observables like $ \rho $\footnote{Let us also remind that extraction of $ \rho $ is inherently model dependent procedure. Unfortunately, no model independent methods are invented by now. Only a kind of the `` proton holography" could help but technically it seems unattainable.}  but rather the difference between $ d\sigma_{\bar{p}p}/dt $ and $ d\sigma_{pp}/dt $ in the vicinity of the `` break(shoulder)"  ($ \bar{p}p $) and dip ($ pp $).}

Coming back to our  criticism of the FOM model , we  in no way claim that a satisfactory $ t\neq  0 $ generalization of Eq.(2) free of the considered above problems and flaws is impossible. 
\textit{Quaerite et invenietis (...fortasse )}. 

Regardless of its model incarnation, the very idea of the Odderon - one of the authors and the most active preacher of which is B. Nicolescu - remains one of the most important quests of modern high energy physics.
 
 \section{Acknowledgements}
Early discussions with Simone Giani and Jan Ka\v{s}par were an inspiring starting point for considerations given above. I am thankful to Evgen Martynov and Basarab Nicolescu for clarifying correspondence on some details of the FMO model.

I am also very much indebted to Anatoliy Samokhin for many fruitful discussions and to the referees and Nikolai Tkachenko for valuable comments and suggestions which allowed to significantly improve the content of this paper.


\begin{thebibliography}{99}



\bibitem{Nic}

L. Lukaszuk and B. Nicolescu,

Lett.al Nuovo Cim.\textbf{8}(1973)405,


\bibitem{Che}
Geoffrey F. Chew and Steven C. Frautschi,

Phys. Rev. \textbf{123}, 1478 

\bibitem{Gri}

V. N. Gribov,

Sov.J.Nucl.Phys. \textbf{17} (1973) 313 

(Yad.Fiz. \textbf{17 }(1973) 603);

Valuable comments to this idea 
taking into account 
the subsequent development can be found in

V. V. Anisovich,

Int.J.Mod.Phys. A\textbf{31} (2016) no.28\&29, 1645010.

\bibitem{Okor}

 V.A. Petrov and V. A. Okorokov,
 
 Int.J.Mod.Phys. \textbf{A 33} (2018) 13, 1850077, 
 
 e-Print: 1802.01559 [hep-ph].

\bibitem {Kho}

R. McNulty, V.A. Khoze, A.D. Martin, M.G. Ryskin,

Eur.Phys.J.C \textbf{80}, 3 (2020) 288;

P. Lebiedowicz, O.Nachtmann, A. Szczurek,

PoS EPS-HEP2019 (2020) 531.




 \bibitem{Ant}

G. Antchev et al. The TOTEM Collaboration,

 Eur. Phys. J. C (2019) \textbf{79 }:785.
 
 \bibitem {Ppe}
 
Vladimir A. Petrov,
      
Eur.Phys.J.C 78 (2018) 3, 221, Eur.Phys.J.C 78 (2018) 5, 414 (erratum);

V.V. Ezhela, V.A. Petrov and  N.P. Tkachenko,

Yad. Fiz. \textbf{84}, 2 (2021)1 (in Russian),

Phys.At.Nucl. \textbf{84}(3),298-313(2021);

https://arxiv.org/abs/2003.03817.

\bibitem{NicMa}

E.Martynov and B. Nicolescu,

a)Phys.Lett.B 778 (2018) 414-418; 

b)Eur.Phys.J. C\textbf{79} (2019) no.6, 461. 

\bibitem {Ni}

E.Martynov and B. Nicolescu,

EPJ Web Conf. \textbf{206} (2019) 06001.



\bibitem{Kh}

V.A. Khoze, A. D. Martin, M.G. Ryskin,

Phys.Lett. B780(2018)

\bibitem {MaNi}

E. Martynov and B. Nicolescu,

Eur. Phys. J. C \textbf{56} (2008) 57.

\bibitem {PDG}
Review of Particle Physics.

P.A. Zyla et al.(Particle Data Group), 

Progr. Theor. Exp. Phys. 2020, 083C01 (2020), p.631.


\bibitem{Dur}

V.A. Khoze, A.D. Martin, M.G. Ryskin, 

Phys. Rev. D97, 034019(2018).

\bibitem{Sel}
 O.V. Selyugin and  J.R. Cudell,
 
 Acta Phys.Polon.Supp. \textbf{12} (2019) 4, 741.



\bibitem{Ro}

 C. Royon( D0 and TOTEM Collaborations),
 
 PoS ICHEP2020, p. 496 (2021),
 
 e-Print:2012.03150 [hep-ex].
 
 \bibitem{Bre}
 A.Breakstone et al.,
 
Phys.Rev.Lett. \textbf{54} (1985) 2180.
  
\end{thebibliography}
\end{document}